\documentclass[12pt]{article}
\usepackage{cite}
\begin{document}
\title{\bf Cosmological parameters from the thermodynamic model of gravity}
\author{{\bf Merab Gogberashvili}\\
Andronikashvili Institute of Physics\\
6 Tamarashvili Street, Tbilisi 0177, Georgia\\
and\\
Javakhishvili State University\\
3 Chavchavadze Avenue, Tbilisi 0179, Georgia\\
{\sl E-mail: gogber@gmail.com}
\\
\\
{\bf Igor Kanatchikov}\\
National Quantum Information Centre of Gda\'nsk\\
ul. W\l. Andersa 27, 81-824 Sopot, Poland \\
{\sl E-mail: kai@fuw.edu.pl}
\date{October 10, 2012}
}
\maketitle
\begin{abstract}
Within our recent thermodynamic model of gravity the dark energy is identified with the energy of collective gravitational interactions of all particles in the universe, which is missing in the standard treatments. For a simple model universe composed of neutral and charged particles of identical mass we estimate the radiation, baryon and dark energy densities and obtain the values which are very close to the current cosmological observations.

\end{abstract}

\vskip 0.5cm


The notions from thermodynamics and statistical physics turned out to be very useful in the field theory context. For instance, thermodynamic arguments have been used in black hole physics \cite{BH-1,BH-2,BH-3}, discovery of Unruh temperature \cite{Unr}, establishment of the AdS/CFT correspondence \cite{AdS}, derivation of the Einstein \cite{Ein} and Maxwell \cite{Wang} equations, and in the recent attempt to interpret the Newtonian gravity as an entropic force \cite{Ver}, and other discussions of the "emergent gravity" \cite{Pad-1, Pad-2,Pad-3}. Besides, the analogies with the classical statistical mechanics and thermodynamics have been underlying some recent discussions of the foundations of quantum mechanics \cite{Wet-1,Wet-2,Wet-3,Wet-4}. However, these relationships are mere analogies so far, since in the context of field theory the physical origin of the microscopic degrees of freedom responsible for the thermodynamic quantities is not yet clear.

As an attempt to address this main question a thermodynamic model of gravity was proposed \cite{Gog-1,Gog-2,Gog-3,Gog-4} in which the universe is considered as the statistical ensemble of all gravitationally interacting particles inside the horizon. According to the model there are no gravitating systems isolated from the world ensemble (which is also quite in the spirit of quantum mechanics) and a more appropriate way to describe gravity is in terms of changes of the local thermodynamic properties in the world ensemble, like temperature or entropy, rather than in terms of the space-time geometry, which we derived as an emerging effective description. It was demonstrated that the model can be compatible with the existing field-theoretical descriptions, as the relativistic and quantum properties are emerging from the properties of the world ensemble. Moreover, the model also allows us to explain the hierarchy problem in particle physics by the fact that our underlying assumption that any gravitational interaction of two particles involves interactions with all particles of the world ensemble effectively weakens the observed strength of gravity by the factor proportional to the number of particles in the ensemble.

In this paper we show that estimations based on the thermodynamic model of gravity also lead to the realistic values of the cosmological parameters of the universe.

In our thermodynamic model \cite{Gog-1,Gog-2,Gog-3,Gog-4} the fundamental constants are collective characteristics of the world ensemble. For instance, the universal constant of the speed of light is connected with the collective gravitational potential of all particles in the universe, $\Phi$, acting on each member of the ensemble:
\begin{equation} \label{Phi}
c^2 = - \Phi = \frac{2GM}{R}~,
\end{equation}
where $G$ is the Newton's constant and $M$ and $R$ are the total mass and the radius of the universe, respectively. This 'universal' potential $\Phi$, and thus $c$, can be regarded as constants, since, according to the cosmological principle, the universe is isotropic and homogeneous at the scale $R$.

The radius of the universe can be defined using the Hubble constant $H$:
\begin{equation} \label{Hubble}
R = \frac cH ~,
\end{equation}
and from (\ref{Phi}), as usual, the total mass of the universe is found:
\begin{equation} \label{M}
M = \frac{c^3}{2 GH} ~.
\end{equation}
Let us emphasize that (\ref{Phi}) is equivalent to the critical density condition in relativistic cosmology:
\begin{equation} \label{rho}
\rho_{c} = \frac{3M}{4\pi R^3} = \frac{3H^2}{8\pi G} ~.
\end{equation}

The assumption (\ref{Phi}) allows us to relate the rest energy of a particle of mass $m$ to its interactions with the whole universe, namely,
\begin{equation} \label{mPhi}
E = mc^2 = - m\Phi~,
\end{equation}
which is equivalent to
\begin{equation} \label{balance}
mc^2 + m\Phi = 0~ .
\end{equation}
The latter equality represents the simplest thermodynamic energy balance equation within the model \cite{Gog-1,Gog-2,Gog-3,Gog-4}. The energy balance conditions mean that the total energy of any object in the universe vanishes, with the gravitational energy of the interaction of any object with the universe assumed to be negative, while all other forms of energy are positive \cite{FMW}. Hence, the total energy of the whole universe vanishes, and thus the universe can emerge without violation of the energy conservation. This is the point of view which appears to be preferable in cosmology \cite{FMW,Haw}.

Let us consider the simplest case of the universe as the ensemble of $N$ identical particles of mass $m$. Since each particle interacts with all other $(N-1)$ particles, and the mean separation of the interacting pairs is $R/2$, the total energy of the ensemble consists of $N(N-1)/2$ terms of magnitude $\approx 2G m^2/R$. Then the energy of a single particle interacting with the total gravitational potential of the universe $\Phi$ is given by:
\begin{equation} \label{E}
E \approx N^2 \frac {Gm^2}{R} ~.
\end{equation}
Therefore, the gravitational mass of the universe is a quadratic function of the number of particles:
\begin{equation} \label{M=N2m}
M_G \approx \frac 12 N^2 m ~.
\end{equation}

It is natural to identify the gravitational energy of collective interactions of all particles with the dark energy of the universe, which, according to the standard $\Lambda$CDM model, is identified with the cosmological constant $\Lambda$. Thus, we assume:
\begin{equation} \label{Edark}
\Omega_\Lambda :=
\frac{M_G}{M} \approx \frac{N^2m}{2M} ~.
\end{equation}
Note that under the above identification the energy balance condition (\ref{balance}) written for all $N$ particles in the universe is equivalent to the $w=-1$ equation of state of the dark energy:
\begin{equation}
\rho + p = 0~,
\end{equation}
where $\rho$ is the dark energy density. In our model the appearance of the "exotic negative pressure $p$" has a natural explanation as the consequence of the negative collective gravitational potential $\Phi$ of the whole universe. Besides, the assumption (\ref{Edark}) also naturally answers the question why the density of dark energy is so close to the critical density (\ref{rho}). The standard cosmology offers no reasonable explanation of this observational fact and the attempt to relate $\Lambda$ to the quantum vacuum fluctuations leads to the value which is 120 orders of magnitude higher than $\rho_c$.

Another universal physical constant is the Planck's action quantum. In our model \cite{Gog-1,Gog-2,Gog-3,Gog-4} it is identified with the characteristic amount of action of a single member of the world ensemble:
\begin{equation} \label{A}
A = - m c \lambda =: - 2\pi\hbar ~,
\end{equation}
were $\lambda$ is a characteristic length of a particle when it can be considered as an oscillator, i.e. its Compton wavelength. The maximal possible length scale of the world ensemble, or the radius of the model universe, is given by
\begin{equation}\label{lambda}
R = \lambda N ~.
\end{equation}

Accordingly, using (\ref{M=N2m}), (\ref{Edark}), (\ref{A}) and (\ref{lambda}) we find the total action of the universe:
\begin{equation} \label{A_U}
A_U := \frac{M c^2}{H} \approx \frac{N^3A}{2\Omega_\Lambda}~,
\end{equation}
and the number of typical particles in it:
\begin{equation} \label{N}
N \approx \left(\frac{2 \Omega_\Lambda A_U}{A} \right)^{1/3} \approx \left(\frac{\Omega_\Lambda M c^2}{\pi \hbar H} \right)^{1/3} \approx 10^{40} ~.
\end{equation}
This number is one of the main parameters of our model and it is known to have appeared in a different context in the Dirac's 'large numbers' hypothesis \cite{large}, and is usually considered as a manifestation of a deep connection between the physics at the subatomic and cosmological scales.

Using the estimation (\ref{N}) and the formulae (\ref{Phi}), (\ref{M=N2m}), (\ref{A}) and (\ref{lambda}), from (\ref{Edark}) we can express the value of the dark energy density in our model universe in terms of the fundamental physical parameters:
\begin{equation}
\Omega_\Lambda = N^3\frac {2\pi \hbar H^2 G}{c^5} \approx 0.72~,
\end{equation}
which is very close to the observed value $\Omega_\Lambda \approx 0.728 \pm 0.015$ \cite{Wmap}.

Now, let us consider a little bit more realistic model universe which also includes both neutral and charged particles of the same mass $m$. The universe as a whole is neutral, i.e. a half of charged particles carries positive charge $+e$ and the other half have negative charge $-e$. The number of charged particles can be roughly identified with the number of baryons in the universe $N_b < N$. A simple combinatorics yields for the gravitational energy of single baryon which interacts with all other particles in the universe the following formula:
\begin{equation} \label{Eb}
E_{b} = (2N_b N - N_b^2)~\frac {Gm^2}{R} \approx 2 N_b N~\frac {Gm^2}{R}~.
\end{equation}
Therefore, according to (\ref{M=N2m}), the total gravitational energy of the baryon component of matter is given by:
\begin{equation} \label{Eb|G}
E_{b|G} = \frac {N^2}{2} E_b \approx N_b N^3~\frac {Gm^2}{R}~.
\end{equation}
It is natural to expect that the ratio (\ref{Edark}) of the gravitational and total energy is valid also for the corresponding contributions of the baryon component:
\begin{equation} \label{E/E}
\frac{E_{b|G}}{E_{b|tot}} \approx \Omega_\Lambda~,
\end{equation}
where $E_{b|tot}$ denotes the total energy of the baryon component of the universe. Then for the baryon density in the universe we obtain the following estimation:
\begin{equation} \label{Omegab}
\Omega_b \approx \frac{E_{b|tot}-E_{b|G}}{Mc^2} \approx \frac{E_{b|G}}{Mc^2}~\frac{(1-\Omega_\Lambda)}{\Omega_\Lambda}~.
\end{equation}

Further, let us estimate the total electromagnetic energy of all $N_b$ charged particles in the model universe (i.e. that of $N_b/2$ interacting pairs). The fact that electric charges have two polarities, while the mass is always positive, leads to basic differences from the previous consideration for the electrically neutral matter. Namely, the universe as a whole is neutral and, in contrast to the gravitational energy, the total electromagnetic, or radiation energy of single baryon consists of $N_b/2$ additive terms, i.e.
\begin{equation}\label{Er}
E_r \approx \frac {N_b}{2} \frac{\alpha \hbar c}{R}~,
\end{equation}
where $\alpha$ is the fine structure constant. Similar to (\ref{Eb|G}) the total gravitational energy of the radiation component of matter is estimated to be
\begin{equation}\label{Er|G}
E_{r|G} \approx N_b N^2\frac{\alpha \hbar c}{4R}~.
\end{equation}
Using this formula and the observed value of the radiation energy density \cite{Wmap}:
\begin{equation}
\Omega_r = \frac {E_{r|G}}{\Omega_\Lambda Mc^2} \approx 4.8 \pm 0.04 \times 10^{-5}~,
\end{equation}
we obtain the estimation of the number of baryons in the universe:
\begin{equation}
N_b \sim 10^{39}~,
\end{equation}
which turns out to be only one order of magnitude less than the estimated total number of particles in our model universe, see (\ref{N}).

Finally, equations (\ref{Eb|G}), (\ref{E/E}), (\ref{Omegab}) and (\ref{Er|G}) yield for the ratio of the radiation and baryon densities in the universe:
\begin{equation} \label{omega/omega}
\frac {\Omega_r}{\Omega_b} \approx \frac{E_{r|G}}{E_{b|G}}~\frac{\Omega_\Lambda}{(1 -\Omega_\Lambda)} \approx \frac{\alpha \hbar c}{4 N G m^2}~\frac{\Omega_\Lambda}{(1-\Omega_\Lambda)}~.
\end{equation}
From (\ref{Phi}) and (\ref{A}) we find:
\begin{equation}
\frac{NGm^2}{\hbar c} = \frac{1}{2\pi \Omega_\Lambda}~,
\end{equation}
whence it follows:
\begin{equation}\label{alpha}
\frac {\Omega_r}{\Omega_b} \approx \frac {\alpha}{8\pi (1-\Omega_\Lambda)} = 1.07 \times 10^{-3}~,
\end{equation}
which is also  very close to the observable value $\Omega_r/\Omega_b \approx 1.09 \pm 0.03 \times 10^{-3}$ \cite{Wmap}.

To conclude, we have considered the energy content of the universe within our recent thermodynamic model of gravity \cite{Gog-1,Gog-2,Gog-3,Gog-4}. The energy of collective gravitational interactions of all particles in the universe is identified with the dark energy. When the existence of charged particles and their contribution to the total gravitational energy is taken into account, it allows us to express some cosmological parameters, namely, the radiation, baryon and dark energy densities of the universe, in terms of the fundamental constants, and to obtain the numerical values of them which are close to the observable ones.



\begin{thebibliography}{99}

\bibitem{BH-1} J.M. Bardeen, B. Carter and S.W. Hawking,
              Comm. Math. Phys. {\bf 31} (1973) 161.

\bibitem{BH-2} J.D. Bekenstein,
              Phys. Rev. {\bf D 7} (1973) 2333.

\bibitem{BH-3} S.W. Hawking,
              Comm. Math. Phys. {\bf 43} (1975) 199.

\bibitem{Unr} W.G. Unruh,
             Phys. Rev. {\bf D 14} (1976) 870.

\bibitem{AdS} J.M. Maldacena,
             Adv. Theor. Math. Phys. {\bf 2} (1998) 231,
             arXiv: hep-th/9711200.

\bibitem{Ein} T. Jacobson,
             Phys. Rev. Lett. {\bf 75} (1995) 1260,
             arXiv: gr-qc/9504004.

\bibitem{Wang} T. Wang,
              Phys. Rev. {\bf 81} (2010) 104045,
              arXiv: 1001.4965 [hep-th].

\bibitem{Ver} E.P. Verlinde,
             JHEP {\bf 1104} (2011) 029,
             arXiv: 1001.0785 [hep-th].

\bibitem{Pad-1} T. Padmanabhan,
               Class. Quant. Grav. {\bf 21} (2004) 4485,
               arXiv: gr-qc/0308070.

\bibitem{Pad-2} T. Padmanabhan,
               Rep. Prog. Phys. {\bf 73} (2010) 046901,
               arXiv: 0911.5004 [gr-qc].

\bibitem{Pad-3} B.L. Hu,
               Int. J. Mod. Phys. {\bf D 20} (2011) 697,
               arXiv: 1010.5837 [gr-qc].

\bibitem{Wet-1} C. Wetterich,
               Ann. Phys. {\bf 325} (2010) 1359,
               arXiv: 1003.3351 [quant-ph].

\bibitem{Wet-2} C. Wetterich,
               J. Phys.: Conf. Ser. {\bf 174} (2009) 012008,
               arXiv: 0811.0927 [quant-ph].

\bibitem{Wet-3} G. Gr\"ossing,
               Physica {\bf A 388} (2009) 811,
               arXiv: 0808.3539 [quant-ph].

\bibitem{Wet-4} G. Gr\"ossing,
               Phys. Lett. {\bf A 372} (2008) 4556,
               arXiv: 0711.4954 [quant-ph].

\bibitem{Gog-1} M. Gogberashvili,
               Eur. Phys. J. {\bf C 63} (2009) 317,
               arXiv: 0807.2439 [gr-qc].

\bibitem{Gog-2} M. Gogberashvili,
               Eur. Phys. J. {\bf C 54} (2008) 671,
               arXiv: 0707.4308 [hep-th].

\bibitem{Gog-3} M. Gogberashvili,
               Int. J. Theor. Phys. {\bf 50} (2011) 2391,
               arXiv: 1008.2544 [gr-qc].

\bibitem{Gog-4} M. Gogberashvili and I. Kanatchikov,
               Int. J. Theor. Phys. {\bf 51} (2012) 985,
               arXiv: 1012.5914 [physics.gen-ph].

\bibitem{FMW} R.P. Feynman, F.B. Morinigo and G. Wagner,
             {\it Feynman lectures on gravitation}
             (Addison-Wesley, Reading 1995).

\bibitem{Haw} S. Hawking,
             {\it A Brief History of Time} (Bantam, Toronto 1988).

\bibitem{large} P.A.M. Dirac,
               Proc. Roy. Soc. London {\bf A 338} (1974) 439.

\bibitem{Wmap} N. Jarosik et. al. (WMAP Collaboration),
              Astrophys. J. Suppl. {\bf 192} (2011) 14,
              arXiv: 1001.4744 [astro-ph.CO].

\end{thebibliography}
\end{document}